\documentstyle[aps,pre,multicol,psfig,12pt]{revtex}
\begin{document}
\tighten
\title{Hydrodynamic Model for Particle Size Segregation in Granular Media}
\author{Leonardo Trujillo$^1$ and Hans J. Herrmann$^{1,2}$}
\address{$^1$P.M.M.H.\footnote{UMR CNRS 7636},
\'Ecole Sup\'erieure de Physique et de Chimie Industrielles,
10, rue Vauquelin, 75231 Paris Cedex 05, France \\
$^2$ICA--1, University of Stuttgart, Pfaffenwaldring 27
D-70569 Stuttgart, Germany}


\date{\today}
\maketitle


\begin{abstract}
We present a hydrodynamic theoretical model for ``Brazil nut" size segregation
in granular materials.
We give analytical solutions for the rise velocity
of a large intruder particle immersed in a medium of monodisperse
fluidized small
particles.
We propose a new mechanism for this particle size-segregation due to buoyant
forces caused by density variations which come from differences in the
local ``granular temperature".
The mobility of the particles is modified by the energy dissipation due
to inelastic collisions and this leads to a  different behavior from
what one would expect for an elastic system.
Using our model we can explain the size ratio dependence of the upward
velocity.
\end{abstract}


\section{Introduction}

The physics of granular materials is a subject
of current interest\cite{Herrmann98}. A granular medium is a system of
many macroscopic heterogeneous particles with dissipative interactions.
One of the outstanding problems is the so--called ``Brazil Nut
effect"\cite{Rosato87}: When a large intruder particle  placed at the
bottom of a vibrated bed tends to the top.
This size segregation  is due to the
nonequilibrium, dissipative nature of  granular media.
Granular materials are handled in many industries.
Many industrial machines that transport granular materials use vertical
vibration to fluidize the material, and the quality of many products is
affected by segregation.
Size segregation is one of the most intriguing phenomena found in
granular physics. A deeper understanding of this effect is therefore interesting
for practical applications, and also represent theoretical challenge.

A series of experiments\cite{Duran93,Knight93,Duran94,Cooke96,Vanel97,Brone97,Shinbrot98}
and computer simulations\cite{Rosato87,Jullien92,Jullien93a,Poschel95,Ohtsuki95,Dippel95,Gallas96,Caglioti98,Shishodia01}
have elucidated different size segregation mechanisms, including vibration frequency and
amplitude\cite{Duran93,Knight93,Duran94,Cooke96,Vanel97,Brone97,Jullien92,Gallas96};
particle size\cite{Duran93,Duran94,Cooke96,Jullien92,Jullien93a,Ohtsuki95,Dippel95,Shishodia01}
and size distribution\cite{Poschel95,Gallas96}; particle shape\cite{Caglioti98};
and other  properties such as density\cite{Shinbrot98,Ohtsuki95,Shishodia01}
and elastic modulus\cite{Gallas96}.

Several possible mechanisms for
size segregation have been proposed. One is segregation in presence of
{\it convection} observed experimentally in three dimensions by
Knight {\it et al.,}\cite{Knight93}, and by Duran {\it al.,}
in two dimensions\cite{Duran94}, under conditions of low amplitude and high acceleration
vibration.
In this case, both intruder and the small particles are driven up along the
middle of the cell, and while the smaller particles are carried down in a
convection roll near the walls the intruder remains trapped on the
top.
In experiments performed by Vanel {\it et al.,} in three
dimensions, they observed two convective regimes separated by a critical frequency\cite{Vanel97}.
The first regime is associated with heaping and the second regime is similar
to the one observed in Ref.\cite{Knight93}. Also, they reported a
{\it nonconvective} regime observing a size dependent rise
velocity.
Employing large molecular dynamics simulations in two dimensions P\"oschel and Herrmann\cite{Poschel95},
and in three dimension Gallas {\it et al.,}\cite{Gallas96}, have
recovered several aspects that are seen in experiments and recognize the
lack of a theoretical description of the exact mechanism driving the
segregation and the role of convection.

Other segregation mechanism is associated to the {\it percolation} of small grains.
Based on a Monte Carlo computer simulation, Rosato {\it et al.,}\cite{Rosato87}
argue that  each cycle of the applied vibration causes all the grains to detach
from the base of the container. Then, the smaller particles
fall relatively freely, while the larger particles require larger voids
to fall downwards. The large grains therefore effectively rise through the bed.
In the context of large--amplitude, low--frequency vertical shaking process (tapping),
Jullien {\it et al.,} predict a critical size ratio below which
segregation does not occur\cite{Jullien92,Jullien93a}.
This provoked some
controversy\cite{Barker93a,Jullien93b,Barker95,Jullien95} and this
threshold may be an artifact of the simulation model
based on the ``steepest descent algorithm"\cite{Jullien92}.
Experiments in Helle Shaw cells\cite{Duran93,Duran94,Cooke96} observed
an intruder size dependent behavior, where the segregation rate increases
with the size ratio between the intruder and the surrounding particles.
Duran {\it et al.,} formulated a geometrical theory for
segregation based on the {\it arching} effect\cite{Duran93}.
They also claim  experimental evidence for a segregation size
threshold\cite{Duran94}.
In this picture the intruder  contributes to the formation of an arch
sustained on small grains on both sides. Between each agitation the
small particles tend to fill the region below the arch. So, at each cycle
the small particles move downward and the intruder effectively rises.
Using a modification of the algorithm
proposed by Rosato {\it et al.}, Dippel and Luding
find a good qualitative agreement with the non--convective and size--dependent
rising\cite{Dippel95}.

In another context, Caglioti {\it et al.,} considered the geometrical
properties of mixtures in the presence of {\it compaction}\cite{Caglioti98}.
They established a relation for the effective mobilities of
different particles in heterogeneous situations.

The effect of the intruder density was studied by Shinbrot and Muzzio\cite{Shinbrot98}.
They observed an oscillating motion of the intruder on the top, which
corresponds to the ``whale effect" predicted by
P\"oschel and Herrmann\cite{Poschel95}.
Also, they observed a reverse buoyancy in shaken granular beds.
Ohtsuki {\it et al.,} performed molecular dynamics simulations in
two dimensions and studied the effects of  intruder size and density on
the height, and found no segregation threshold\cite{Ohtsuki95}.
Recently Shishoda and Wassgren performed two dimensional simulations
to model segregation in vibrofluidized beds\cite{Shishodia01}. They reported
an  height dependence with the density ratio between the intruder and
the surrounding particles. In their model the intruder position results from
a balance between the granular pressure  ({\it buoyant} force) within the bed and the
intruder weight. Their approach is in some sense similar to the model
that we propose in this article.

Subject to an external force, granular materials locally perform random
motions as a result of collisions between grains, much like the
molecules in a  gas.
This picture has inspired several authors
to use kinetic theories
to derive continuum equations for the granular flow--field
variables\cite{Haff83,Jenkins85,Savage88,Jenkins88,Campbell91,Jenkins98,Kumaran98,Garzo99}.
Some of these
theories have been generalized to multicomponent mixtures of
grains\cite{Jenkins89,Zamankhan95,Arnarson98,Willits99}.
For different size particles in the presence of a temperature gradient,
Arnarson and Willits, found that larger, denser particles tend to be more
concentrated in cooler regions\cite{Arnarson98}. This result was confirmed
by numerical simulations\cite{McNamara00,Henrique01}. However, this mechanism
of segregation is a natural consequence of the imposed gradient of temperature
and its not related to the nature of the grains\cite{Henrique01}.

In this article we address the problem
of size segregation using a kinetic theory approach in two and three
dimensions ($D=2,3$).
We consider the case of an intruder particle immersed in a granular bed.
We propose a segregation mechanism based on the difference of
densities between different regions of the system,
which give origin to a buoyant force that acts on the intruder.
The difference of densities is caused
by the difference between the mean kinetic energy among the region around the intruder
and the medium without intruder.
The dissipative nature of the collisions between the particles of a
granular media is  responsible for this mean energy difference,
and modifies the mobility of the particles.

The plan of this article is as follows. In Sec. II we derive
a continuum formulation for the granular fluid, and introduce
the definition of the ``granular temperature".
In Sec. III we propose an analytic method to estimate the local
temperature in the system.
In Sec. IV we introduce the coefficient of thermal expansion.
In Sec. V explicit solutions of the time dependence of  height and velocity
of the large particle are calculated.
We can explain the size ratio dependence of the rise velocity
and address the issue of the critical size ratio to segregation.
To validate our arguments  we make comparisons with previous experimental
data.

\section{Continuum formulation}

We consider an intruder particle of mass $m_{I}$ and radius $r_{I}$ immersed in a
granular bed. The granular bed is formed of $N$ monodisperse particles
of mass $m_F$ and radius $r_F$. The particles are modeled by
inelastic hard disks ($D=2$) or spheres ($D=3$) in a $D$--dimensional
volume $V=L^D$ of size $L$. The size ratio is denoted $\phi = r_I/r_F$.
The particles interact via binary encounters. The inelasticity is
specified by a restitution coefficient $e \leq 1$.
We assume this restitution coefficient to be a constant, independent on
the impact velocity and the same for the fluid particles and the intruder.
The post collisional velocities ${\bf v}'$ are given in terms of the
pre--collisional velocities ${\bf v}$ by
\begin{equation}
{\bf v}_{1,2}'={\bf v}_{1,2}\mp\frac{m_{red}(1+e)}{m_{1,2}}[({\bf v}_{1}-{\bf v}_{2})\cdot{\bf \hat{n}}]{\bf \hat{n}},
\label{Vel}
\end{equation}
where the labels $1$ or $2$ specify the particle, ${\bf \hat{n}}$ is the unit
vector normal to the tangential contact plane pointing
from $1$ to $2$ at the contact time, and the reduced mass
$m_{red}=m_1 m_2/(m_1 + m_2)$.
To calculate the dissipated energy we
consider that energy is dissipated only by collisions between pairs of
grains.
In a binary collision the energy dissipated is proportional to
$\Delta E= -m_{red}(1-e^2)v^2/2$, where $v$ is the mean velocity
of the particles.

In this work we use a generalized notion of temperature.
In a vibrofluidized granular material a ``granular temperature'' $T_g$ can be
defined to describe the random motion of the grains and is the
responsible for the pressure, and the transport of momentum and energy in the
system\cite{Campbell91}.
The granular temperature $T_g$ is defined proportional to the mean kinetic energy
$E$ associated to the velocity of each particle
\begin{equation}
\frac{D}{2} T_g = \frac{E}{N}=\frac{1}{N}\sum_{i=1}^{N} \left( \frac{1}{2}m_i v_i^2 \right).
\end{equation}

We expect a continuum limit to hold for $N \gg 1$, when the small particles
may be considered as forming a granular fluid.
In order to develop an analytic study, we assume that the uniformly heated
granular fluid
can be described by the standard hydrodynamic equations
derived from kinetic theories for granular systems\cite{Haff83}.
In this study, we focus on a steady state with no macroscopic flow.

The balance equation for the energy is
\begin{equation}
{\bf \nabla}\cdot {\bf q}= - \gamma,
\label{energy}
\end{equation}
where ${\bf q}$ is the flux of energy and $\gamma$ is the
average rate of dissipated energy due to the inelastic nature of the
particles collisions.
The constitutive relation for the flux of  energy,
\begin{equation}
{\bf q}=-\kappa{\bf \nabla}T_g,
\label{flux}
\end{equation}
defines the thermal conductivity $\kappa$.
Consequently, we have
\begin{equation}
{\bf \nabla}\cdot \left( \kappa {\bf \nabla}T_g \right)=\gamma.
\label{en-bal}
\end{equation}

A uniformly fluidized state can be realized when the granular material is vibrated
in the vertical direction, typically as $z_0(t)=A_0\sin (\omega_0 t)$,
with the amplitude $A_0$ and the frequency $\omega_0=2\pi f$, so that
one can define a typical velocity $u_0=A_0\omega_0$. In the experiments the
excitation is described by the dimensionless  amplitude
$\Gamma_0=A_0\omega_0/g$, where $g$ is the gravitational acceleration.
As a first approximation the effect of the external force experienced by
the particles due to the gravitational field is neglected in
the description of the granular flow.
Experimentally this corresponds to the regime $\Gamma_0\gg 1$.
So, the momentum balance, in the
steady state, implies that the pressure $p$ is constant throughout the
system.

The hydrodynamic equations close with the state equation,
the collisional dissipation $\gamma$ and the transport coefficients for
a granular medium.
In the limit $N\gg 1$ the constitutive relations
are determined as function of the properties of the small grains.
The transport coefficients are assumed to be given by the
Enskog theory for dense gases in the limit of small inelasticity.

The total pressure should be essentially equal to that of the small
particles, the contribution of the intruder being negligible, since
$N\gg 1$. For a dense system the pressure is related to the density by the
virial equation of state, which in the case of inelastic particles
is\cite{Jenkins85,Garzo99}
\begin{equation}
p=\frac{1+e}{2}nT_g\left[ 1+\frac{\Omega_D}{2D} n g_0 (2r_F)^D \right],
\label{eqstate}
\end{equation}
where $n=N/V$ is the number density
of small grains, $\Omega_D=2\pi^{D/2}/\Gamma(D/2)$ is the surface area of a
$D$--dimensional unit sphere,
$\nu $ is the area (volume) fraction $\nu = \Omega_Dnr_F^D/D$,
and $g_0$ is the pair correlation function
for two fluid particles. In two dimensions the pair correlation function
is\cite{Verlet82}
\begin{equation}
g_0 = \frac{\left( 1 - \frac{7}{16} \nu\right) }{\left( 1-\nu \right)^2},
\label{Verlet}
\end{equation}
with  the area fraction $\nu=n\pi r_F^2$. In three dimensions the pair
correlation function is\cite{Carnahan69}
\begin{equation}
g_0=\frac{\left( 2-\nu \right) }{2\left( 1-\nu\right)^3},
\label{Carnahan}
\end{equation}
with the volume fraction $\nu = 4\pi n r_F^3/3$.

The state--dependent thermal conductivity possesses the general form\cite{Haff83}
\begin{equation}
\kappa = \kappa_0\sqrt{T_g},
\label{conductivity}
\end{equation}
where the prefactor $\kappa_0$ is a function of the fluid
particle properties, and can be calculated using a
Chapman--Enskog procedure through the solution of Enskog transport
equation\cite{Jenkins85,Jenkins98,Garzo99,Chapman70}.
The explicit expressions of these prefactors are given in
Appendix A.

To estimate the collisional dissipation rate $\gamma$ we consider the loss
of average kinetic energy per collision and per unit time. In a binary
collision the kinetic energy dissipated can be expressed in terms of the
granular temperature as $\Delta T_g = -(1-e^2) T_g /2$.
For the fluid particles, the average collision frequency $\omega_F$ is
proportional to $\omega_F \sim \sqrt{T_g}$, and we assume that it is given
by the Enskog collision frequency\cite{Chapman70}
\begin{equation}
\omega_F = \frac{\Omega_D}{\sqrt{2\pi}} n g_0(2r_F)^{D-1}\left( \frac{2}{m_F} \right)^{1/2}T_g^{1/2}.
\label{freqF}
\end{equation}
This form for the frequency of collisions is justified for a granular medium.
This is a consequence that the average spacing
between nearest neighbor $s$ is supposed to be less than the grain diameter $(s\ll 2r_F)$\cite{Haff83}.
Multiplying $\Delta T_g$ by the collision rate $\omega_F$ and the number density $n=N/V$,
we obtain the collisional dissipation rate $\gamma_F$ for the fluid particles
\begin{equation}
\gamma_F = \frac{\Omega_D}{2\sqrt{2\pi}}(1-e^2) n^2 g_0 (2r_F)^{D-1}\left( \frac{2}{m_F} \right)^{1/2}T_g^{3/2}.
\label{gammaF}
\end{equation}

In order to simplify the mathematical notation let us express $\gamma_F$ as
\begin{equation}
\gamma_F = \xi_F T_g^{3/2},
\end{equation}
where the dissipation factor $\xi_F$  contains the prefactors which multiply $T_g^{3/2}$
in Eq.(\ref{gammaF}), this is
\begin{equation}
\xi_F \equiv \frac{\Omega_D}{2\sqrt{2\pi}}(1-e^2) n^2 g_0 (2r_F)^{D-1}\left( \frac{2}{m_F} \right)^{1/2}.
\end{equation}

To understand the essential features of the intruder's presence in the granular medium, it is
adequate to adopt a simplified point of view. If the mean velocity of the fluid particles is
$u$, the flux of fluid particles which strikes the intruder's surface can be estimate as
$n u$. Multiplying this flux by the area of the intruder $\Omega_D r_I^{(D-1)}$, we can
calculate the number of fluid particles which strike the surface of the intruder per unit time,
and written in terms of the granular temperature we have
\begin{equation}
\omega_I = \frac{\Omega_D}{\sqrt{2\pi}}ng_0 r_I^{D-1}\left( \frac{m_I + m_F}{m_I m_F} \right)^{1/2}T_g^{1/2}.
\label{freqI}
\end{equation}
So, the local density of kinetic energy dissipated in the region near the intruder is
\begin{equation}
\gamma_I = \frac{\Omega_D}{2\sqrt{2\pi}}(1-e^2) \frac{n}{V} g_0(r_F+ r_I)^{D-1} \left( \frac{m_I + m_F}{m_I m_F} \right)^{1/2}T_g^{3/2}.
\label{gammaI}
\end{equation}

In the simplified form Eq.(\ref{gammaI}) can be expressed as
\begin{equation}
\gamma_I = \xi_I T_g^{3/2},
\end{equation}
where the dissipation factor $\xi_I$ is defined as
\begin{equation}
\xi_I\equiv \frac{\Omega_D}{2\sqrt{2\pi}}(1-e^2)\frac{n}{V}g_0(r_F+ r_I)^{D-1} \left( \frac{m_I + m_F}{m_I m_F} \right)^{1/2}.
\end{equation}
%

\section{Local temperature difference}

The intruder's presence
modifies the local temperature of the system due to the collisions that happen
at its surface.
The number of collisions on the surface increases with
the size of the particle, but the local density of dissipated energy
diminishes.
From Eq.(\ref{en-bal}) we can calculate within a sphere of radius $r_0$
the value of the temperature in the
granular fluid in presence of the intruder
 and compare it with the
temperature in the granular fluid without intruder,
we will denote these temperatures
$T_{1}$ and $T_2$ respectively (see Fig.(1)).
This  is a simple method to estimate the temperature difference
between a region with intruder and a region without intruder
$\Delta T_g =  T_1-T_2$.
\begin{figure}[t]
\centerline{\psfig{file=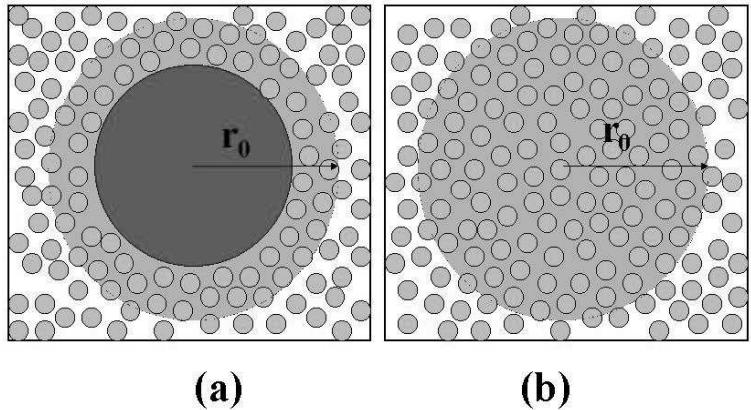,width=10.0cm}}
\caption{Schematic representation of the regions used to  calculate  the granular temperature.
(a) Region around the intruder within a sphere of radius $r_0$ and
(b) region without intruder.
\label{fig1}}
\end{figure}

Let us concentrate on solutions with radial symmetry. The solutions
of Eq.(\ref{en-bal}), for an arbitrary dimension $D$, satisfy the equation
\begin{equation}
\frac{1}{r^{D-1}}\frac{d}{dr}\left( r^{D-1} \kappa_0 T_g^{1/2}\frac{dT_g}{dr} \right)=\xi T_g^{3/2}.
\label{ec1}
\end{equation}
This nonlinear differential equation can be simplified by the fact
that the pressure is considered constant throughout the system
and remembering that $p\sim T_g$.
So, linearizing Eq.(\ref{ec1}) the resulting equation may
be written in terms of $w\equiv T_g^{1/2}$,
\begin{equation}
\frac{d^2 w}{dr^2}+\frac{D-1}{r}\frac{dw}{dr}=\lambda^2 w,
\label{eqw}
\end{equation}
where
\begin{equation}
\lambda^2 \equiv \frac{\xi}{2\kappa_0}.
\end{equation}

The collisional dissipation rate can be decomposed in two parts. We propose this
decomposition supposing that the energy dissipation around the intruder is
dominated by the collisions between the small grains and the intruder, then
the dissipation rate in this region is given by Eq.(\ref{gammaI}). In the rest of the
system the dissipation rate is dominated by the collisions between small grains only. In
this case the dissipation is given by Eq.(\ref{gammaF}).

First, let us consider the ``inhomogeneous case'' when the intruder is localized in the center
of the system ($r=0$), see Fig.(1a). The dissipation factor $\xi$ can be decomposed in
two parts: $\xi=\xi_{I}$ for the region near the intruder $(r=r_I)$, and
$\xi=\xi_{F}$ for the region $(r_I<r\leq r_0)$, where $r_0$ is the radius of the considered
region.

For the inhomogeneous case we express Eq.(\ref{eqw}) as
\begin{equation}
\frac{d^2 w}{dr^2}+\frac{D-1}{r}\frac{dw}{dr} = \left\{
\begin{array}{ll}
\lambda_I^2w & \mbox{for $0<r\leq r_I$}, \\
\\
\lambda_F^2w & \mbox{for $r_I<r\leq r_0$},
\end{array}
\right.
\label{ind}
\end{equation}
where, $\lambda_{I}^{2}\equiv\xi_I/2\kappa_0$ and
$\lambda_{F}^{2}\equiv\xi_F/2\kappa_0$. The solution of Eq.(\ref{ind})
is determined by the boundary conditions imposed upon the system. As boundary
condition we suppose that the system is
enclosed by an external surface of radius $r_0$ at temperature $T_g(r_0)=T_0$
(respectively, $w(r_0)=w_0$).

Let us denote $T_{1-}(r)$ the granular temperature for the region
$(0<r\leq r_I)$, and $T_{1+}(r)$ the granular temperature for the
region $(r_I< r \leq r_0)$ (respectively, $w_-(r)\equiv \sqrt{T_{1-}(r)}$
and $w_+(r)\equiv\sqrt{T_{1+}(r)}$). The intruder's presence imposes internal
boundary conditions. On the inner surface, the temperature should satisfy
\begin{equation}
\left. w_-(r)\right|_{r=r_I} = \left. w_+(r)\right|_{r=r_I}.
\label{Bound1}
\end{equation}

The flux of energy also imposes another internal boundary condition. If we suppose
the flux of energy  continuous on the inner surface, from Eq.(\ref{flux}) the
granular temperature should satisfy
\begin{equation}
\left. \frac{dw_-(r)}{dr}\right|_{r=r_I} = \left. \frac{dw_+(r)}{dr}\right|_{r=r_I}.
\label{Bound2}
\end{equation}

\subsection{Solution for 2D}

The solutions to Eq.(\ref{ind}) for $D=2$ are a linear combination of the modified Bessel function
of order zero  $w_1(r)=\{I_0(\lambda r),K_0(\lambda r)\}$.
The general solution is
\begin{equation}
w_-(r) = A_- I_0(\lambda_I r) + B_-K_0(\lambda_I r) \qquad \mbox{ for $0<r\leq r_I$},
\label{SolGen1}
\end{equation}
and
\begin{equation}
w_+(r) = A_+ I_0(\lambda_F r) + B_+K_0(\lambda_F r) \qquad \mbox{ for $r_I<r\leq r_0$},
\label{SolGen2}
\end{equation}
where $A_-$, $A_+$, $B_-$ and $B_+$ are constants that must be determined from
the boundary conditions.

The function $K_0(\lambda r)$ diverges when $r\rightarrow 0$, then
\begin{equation}
B_- = 0.
\end{equation}

When $r=r_0$ the Eq.(\ref{SolGen2}) should satisfy the boundary condition
\begin{equation}
\left. w_+(r)\right|_{r=r_0} = w_0,
\end{equation}
this is,
\begin{equation}
A_+ I_0 (\lambda_F r_0) + B_+ K_0(\lambda_F r_0) = w_0.
\label{cinco}
\end{equation}

On the inner surface the boundary condition (\ref{Bound1}) $w_-(r_I) = w_+(r_I)$ leads to
\begin{eqnarray}
A_- I_0(\lambda_I r_I) & = & A_+ I_0(\lambda_F r_I) + B_+ K_0(\lambda_F r_I)\nonumber \\
\Longrightarrow A_- & = & A_+ \frac{I_0(\lambda_F r_I)}{I_0(\lambda_I r_I)} + B_+\frac{ K_0(\lambda_F r_I)}{I_0(\lambda_I r_I)}.
\label{seis}
\end{eqnarray}

The inner boundary condition (\ref{Bound2}) leads to
\begin{eqnarray}
A_-\lambda_I I_1(\lambda_I r_I) & = & A_+\lambda_F I_1(\lambda_F r_I) - B_+ \lambda_F K_1(\lambda_Fr_I) \nonumber \\
\Longrightarrow A_- & = & \left( \frac{\lambda_F}{\lambda_I} \right)\left[ A_+\frac{I_1(\lambda_F r_I)}{I_1(\lambda_I r_I)} - B_+\frac{K_1(\lambda_F r_I)}{I_1(\lambda_F r_I)}\right].
\label{siete}
\end{eqnarray}

Equating Eqs.(\ref{seis}) and (\ref{siete}) we find
\begin{eqnarray}
\frac{A_+}{B_+} & = &  \frac{\lambda_FI_0(\lambda_Ir_I)K_1(\lambda_Fr_I)+\lambda_II_1(\lambda_Ir_I)K_0(\lambda_Fr_I)}{\lambda_FI_0(\lambda_Ir_I)I_1(\lambda_Fr_I) - \lambda_I I_1(\lambda_Ir_I)I_0(\lambda_Fr_I)}, 
\nonumber
\\
 & \equiv & \Theta_{AB}
\label{theta}
\end{eqnarray}

From Eqs. (\ref{cinco}) and (\ref{theta}) the constant $B_+$ should be
\begin{equation}
B_+ = \frac{w_0}{\Theta_{AB}I_0(\lambda_Fr_0)+K_0(\lambda_Fr_0)}.
\label{B+}
\end{equation}

Substituting Eq.(\ref{B+}) into (\ref{theta}) we have
\begin{equation}
A_+ = \frac{w_0\Theta_{AB}}{\Theta_{AB}I_0(\lambda_Fr_0)+K_0(\lambda_Fr_0)}.
\label{A+}
\end{equation}

Substituting Eqs.(\ref{B+}) and (\ref{A+}) into (\ref{seis}) we have
\begin{equation}
A_- = \frac{w_0}{\Theta_{AB}I_0(\lambda_Fr_0)+K_0(\lambda_Fr_0)} \left[ \Theta_{AB}\frac{I_0(\lambda_Fr_I)}{I_0(\lambda_Ir_I)}  + \frac{K_0(\lambda_Fr_I)}{I_0(\lambda_Ir_I)}\right].
\label{A-}
\end{equation}

The granular temperature in the inhomogeneous case is
\begin{equation}
T_1(r)=\left\{
\begin{array}{ll}
(A_-I_0(\lambda_Ir))^2 & \mbox{for $0<r\leq r_I$},\\
\\
(A_+I_0(\lambda_F r)+ B_+K_0(\lambda_F r))^2 & \mbox{for $r_I<r\leq r_0$},
\end{array}
\right.
\label{Temp12D}
\end{equation}
where the constant $A_-$, $A_+$ and $B_+$ are given by the Eqs.(\ref{A-}), (\ref{A+}) and
(\ref{B+}), respectively.

In the ``homogeneous case", see Fig.(1b), the prefactor $\lambda_I =0$. Then the granular
temperature $T_2(r)$ is
\begin{equation}
T_2(r)=\left(\frac{I_0(\lambda_F r)}{I_0(\lambda_F r_0)}\right)^2 T_0. 
\label{Temp22D}
\end{equation}

Now we are interested in determining the temperature difference $\Delta T_g$ between case $1$ and $2$ in the
granular fluid. For this we calculate the granular temperatures at $r=0$.
When $r\rightarrow 0$ the modified Bessel function of zero order tends to $1$\cite{Abramowitz65}.
So, in Eq.(\ref{Temp12D}) and (\ref{Temp22D}) in the limit $r\rightarrow 0$ we find that
\begin{equation}
\begin{array}{ll}
(I_0(\lambda_I r))^2 \sim 1,\\
(I_0(\lambda_F r))^2 \sim 1.
\end{array}
\end{equation}
Then, the temperature difference is
\begin{equation}
\Delta T_g = \left[\left( \frac{1}{\Theta_{AB}I_0(\lambda_Fr_0)+K_0(\lambda_Fr_0)} \left[ \Theta_{AB}\frac{I_0(\lambda_Fr_I)}{I_0(\lambda_Ir_I)}  + \frac{K_0(\lambda_Fr_I)}{I_0(\lambda_Ir_I)}\right]  \right)^2 - \left(\frac{1}{I_0(\lambda_F r_0)}\right)^2 \right] T_0
\label{deltaT1}
\end{equation}
in two dimensions.

\subsection{Solution for 3D}

When $D=3$, the solution of Eq.(\ref{ind}) is given in terms of the
spherical modified Bessel functions of zero order
$w_1(r)=\{i_0(\lambda r)=\sinh(\lambda r)/\lambda r,k_0(\lambda r)=e^{-\lambda r}/\lambda r\}$.
The general solution in this case is
\begin{equation}
w_-(r) = A_- i_0(\lambda_I r) + B_-k_0(\lambda_I r) \qquad \mbox{ for $0<r\leq r_I$},
\label{1}
\end{equation}
and
\begin{equation}
w_+(r) = A_+ i_0(\lambda_F r) + B_+k_0(\lambda_F r) \qquad \mbox{ for $r_I<r\leq r_0$}.
\label{2}
\end{equation}

The function $k_0(\lambda r)$ diverges when $r\rightarrow 0$, then
\begin{equation}
B_- = 0.
\end{equation}

The constants $A_-$, $A_+$ and $B_+$, are calculated from the boundary conditions in a 
similar way as before. 
\begin{eqnarray}
A_- & = & \frac{w_0}{\Theta_{AB}i_0(\lambda_Fr_0)+k_0(\lambda_Fr_0)} \left[ \Theta_{AB}\frac{i_0(\lambda_Fr_I)}{i_0(\lambda_Ir_I)}  + \frac{k_0(\lambda_Fr_I)}{i_0(\lambda_Ir_I)}\right],
\label{A-3}
\\
A_+ & = & \frac{w_0\Theta_{AB}}{\Theta_{AB}i_0(\lambda_Fr_0)+k_0(\lambda_Fr_0)},
\label{A+3}
\\
B_+ & = & \frac{w_0}{\Theta_{AB}i_0(\lambda_Fr_0)+k_0(\lambda_Fr_0)},
\label{B+3}
\end{eqnarray}
where in this case the factor $\Theta_{AB}$ is

\begin{equation}
\Theta_{AB} =   \frac{\lambda_Fi_0(\lambda_Ir_I)k_1(\lambda_Fr_I)+\lambda_I i_1(\lambda_Ir_I)k_0(\lambda_Fr_I)}{\lambda_F i_0(\lambda_Ir_I)i_1(\lambda_Fr_I) - \lambda_I i_1(\lambda_Ir_I)i_0(\lambda_Fr_I)}.
\label{theta3}
\end{equation}

The granular temperature in the inhomogeneous case in 3D is
\begin{equation}
T_1(r)=\left\{ 
\begin{array}{ll}
(A_-i_0(\lambda_Ir))^2 & \mbox{for $0<r\leq r_I$},\\
\\
(A_+i_0(\lambda_F r)+ B_+k_0(\lambda_F r))^2 & \mbox{for $r_I<r\leq r_0$},
\end{array}
\right.
\label{Temp13D}
\end{equation}
where the constant $A_-$, $A_+$ and $B_+$ are given by the Eqs.(\ref{A-3}), (\ref{A+3}) and
(\ref{B+3}).

In the ``homogeneous case" the prefactor $\lambda_I =0$. Then the granular
temperature $T_2(r)$ is
\begin{equation}
T_2(r)=\left(\frac{i_0(\lambda_F r)}{i_0(\lambda_F r_0)}\right)^2 T_0.
\label{Temp23D}
\end{equation}

Again the temperature difference $\Delta T_g$ is calculate at $r=0$
between case $1$ and $2$.
When $r\rightarrow 0$ the spherical Bessel function of zero order tends to $1$\cite{Abramowitz65}, then
\begin{equation}
\begin{array}{ll}
\left( \frac{\sinh(\lambda_I r)}{\lambda_I r} \right)^2 \sim 1,\\
\left( \frac{\sinh(\lambda_F r)}{\lambda_F r} \right)^2 \sim 1.
\end{array}
\end{equation}
Then, the temperature difference is
\begin{equation}
\Delta T_g = \left[\left( \frac{1}{\Theta_{AB}i_0(\lambda_Fr_0)+k_0(\lambda_Fr_0)} \left[ \Theta_{AB}\frac{i_0(\lambda_Fr_I)}{i_0(\lambda_Ir_I)}  + \frac{k_0(\lambda_Fr_I)}{i_0(\lambda_Ir_I)}\right]  \right)^2 - \left(\frac{1}{i_0(\lambda_F r_0)}\right)^2 \right] T_0
\label{deltaT2}
\end{equation}
in three dimensions.

\subsection{Energy equipartition breakdown}

Let us define the temperature ratio $\tau\equiv T_1(0)/T_2(0)$. In two dimension we have
\begin{equation}
\tau =  \left( \frac{I_0(\lambda_F r_0)}{\Theta_{AB}I_0(\lambda_Fr_0)+K_0(\lambda_Fr_0)} \left[ \Theta_{AB}\frac{I_0(\lambda_Fr_I)}{I_0(\lambda_Ir_I)}  + \frac{K_0(\lambda_Fr_I)}{I_0(\lambda_Ir_I)}\right]  \right)^2,
\end{equation}
and for three dimensions,
\begin{equation}
\tau = \left( \frac{i_0(\lambda_F r_0)}{\Theta_{AB}i_0(\lambda_Fr_0)+k_0(\lambda_Fr_0)} \left[ \Theta_{AB}\frac{i_0(\lambda_Fr_I)}{i_0(\lambda_Ir_I)}  + \frac{k_0(\lambda_Fr_I)}{i_0(\lambda_Ir_I)}\right]  \right)^2,
\end{equation}
since $\lambda_F > \lambda_I$ we can verify that $T_1(0) > T_2(0)$, this means $\tau > 1$. So, the
temperatures ratio between the region with intruder and the region without intruder are different.
In our model this lack of equipartition is due to a difference between the collisional dissipation rate
related to the particle sizes. In the elastic limit $e\rightarrow 1$ the energy equipartition is
restored $\tau \rightarrow 1$. In Fig. 2, we present the qualitative behavior of $\tau$ with
the size ratio $\phi = r_I/r_F$, for different values of the coefficient $e$. The granular
temperature difference increases with $\phi$ and depends on $e$. We can see that $\tau$ is
nearly constant and very close to unity when $e=0.99$.

\begin{figure}[t]
\centerline{\psfig{file=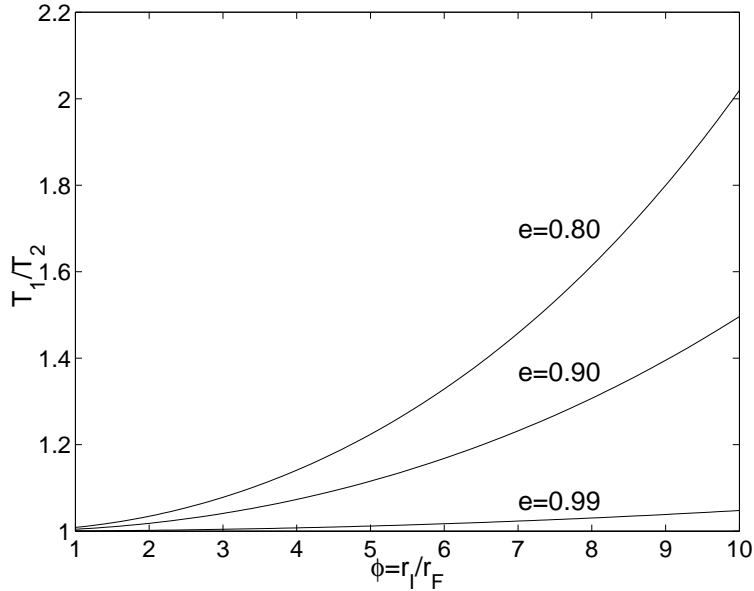,width=10.0cm}}
\caption{Ratio $\tau=T_1(0)/T_2(0)$ of the granular temperatures, showing non--equipartition
of energy ($\tau\neq 1$) for different values of the coefficient of restitution $e$.
\label{fig2}}
\end{figure}

Recently, this quantity was directly measured in experiments performed by Feitosa and Menon\cite{Feitosa01}.
They observed that energy equipartition does not generally hold for a binary vibrated granular system.
They reported that the ratio of granular temperatures depends on the ratio of particle mass densities.
Also in fluidized binary granular mixtures
the breakdown of the energy equipartition was observed experimentally\cite{Ippolito95} and described
theoretically in the framework of the kinetic theory\cite{Huilin01}

\section{Thermal expansion}

The definition of the granular temperature allows us to establish a statistical mechanics description
of the granular medium. The change of mean energy of the system is  basically due to a
mechanical interaction with their external parameters (e.g., the amplitude $A_0$ and the frequency
$\omega_0=2\pi f$ of vibration, the volume of the system $V$, and the pressure $p$).
Theoretically
we can derive the energy relaxation to the steady state for a driven granular medium\cite{Sunthar99}.
The system increases its energy as a result of external driving while its decreases its energy by
dissipation.
The work $W$ done to change the volume of the system from $V$ to a certain quantity $V + dV$
is equal to the change of
its mean energy and its related to the mean pressure and volume by $dW=pdV +Vdp$.
From the definition
of granular temperature, the change of the granular temperature depends on the mean kinetic
energy of the particles. A volume change $dV$ is related to a temperature change $dT_g$ by
the equation of state (\ref{eqstate}).

We can express $V$ as a function of $T_g$ and $p$, $V=V(T_g,p)$. Thus given infinitesimal
changes in $T_g$ and $p$, we can write
\begin{equation}
\begin{array}{ll}
dV & = \left( \frac{\partial V}{\partial T_g} \right )_p dT_g + \left( \frac{\partial V}{\partial p} \right )_{T_g} dp,\\
& = \alpha V dT_g - k_p V dp,
\end{array}
\label{Vol}
\end{equation}
where $\alpha$ is the thermal expansion coefficient defined as
\begin{equation}
\alpha \equiv \frac{1}{V}\left( \frac{\partial V}{\partial T_g}\right)_p =-\frac{1}{n}\left( \frac{\partial n}{\partial T_g}\right)_{p,N},
\label{alfa}
\end{equation}
and $k_p$ is the ``isothermal compressibility" defined as
\begin{equation}
k_p \equiv -\frac{1}{V}\left( \frac{\partial V}{\partial p}\right)_{T_g} =\frac{1}{n}\left( \frac{\partial n}{\partial p}\right)_{T_g}.
\end{equation}

If in a first approximation we neglect the variations of the coefficients $\alpha$ and $k_p$, we
can integrate Eq.(\ref{Vol}) and find
\begin{equation}
\begin{array}{ll}
V(T_g,p) & =V_0\exp \left[ \alpha \Delta T_g - k_p \Delta p \right],\\
& \approx V_0 \left[1 + \alpha \Delta T_g - k_p \Delta p \right].
\end{array}
\label{vol}
\end{equation}

From the temperature difference, Eqs.(\ref{deltaT1}) and (\ref{deltaT2}),
we can conclude that we also have a change in the {\it effective density}
($\rho = nm_I$) of the
granular fluid between case 1 and 2, through the thermal expansion of volume produced by the difference of
temperatures, Eq.(\ref{vol}), thus
\begin{equation}
\tilde \rho = \rho (1-\alpha \Delta T_g).
\end{equation}

The thermal expansion coefficient can be derived from the equation of state (\ref{eqstate}) and
 definition (\ref{alfa}).
The general form of the coefficient of thermal expansion is
\begin{equation}
\alpha = \frac{1}{T_0} C(\nu),
\end{equation}
where $C(\nu)$ is a correction due to the density of the system. In the dilute limit
$\nu \rightarrow 0$ and $C(\nu)\rightarrow 1$, and the above expression tends to the
expected value for a classical gas $\alpha = 1/T_0$. The explicit form of $C(\nu)$ is
given in Appendix B.

\section{Segregation forces}

Now we propose that this density difference leads to an
{\it effective buoyancy force} ${\bf f}_{b}$, similar to the Archimedean force
\begin{equation}
{\bf f}_{b} = \Delta \rho V_I {\bf g},
\label{buoyancy}
\end{equation}
where $\Delta \rho = -\alpha\rho\Delta T_g$, $V_I=\frac{\Omega_D}{D}r_I^D$
is the $D$--dimensional volume of the intruder and ${\bf g}$ is the gravity field.
The intruder also experiences a viscous drag of the granular fluid. The drag force
${\bf f}_d$ is considered to be linear in the velocity of segregation ${\bf u}(t)$, and
is like the Stokes' drag force
\begin{equation}
{\bf f}_d = -6\pi\mu r_I {\bf u}(t),
\label{drag}
\end{equation}
where $\mu$ is the coefficient of viscosity of the granular fluid
The state--dependent viscosity  posesses the general form\cite{Haff83}
\begin{equation}
\mu = \mu_0\sqrt{T_g},
\label{viscosity}
\end{equation}
where the prefactor $\mu_0$ is a function of the fluid
particle properties, and can be calculated using a
Chapman--Enskog procedure for the solution of Enskog transport
equation.
The explicit expressions of these prefactors are given in Appendix A.

Equations (\ref{buoyancy}) and (\ref{drag}) express the acting forces in the
segregation process
\begin{equation}
{\bf f}_{seg}={\bf f}_{b} + {\bf f}_{d}.
\end{equation}
Therefore,
the equation of motion that governs the segregation process is
\begin{equation}
\frac{\Omega_D}{D}r_I^D\rho \frac{d {\bf u}(t)}{dt} = -\frac{\Omega_D}{D}r_I^D\alpha \rho \Delta T_g {\bf g} -6\pi \mu r_I {\bf u}(t).
\label{Eqmov}
\end{equation}

Now we suppose the granular system contained between two large parallel plates
perpendicular to the gravitational field. We take the reference frame positive
in the upward vertical direction.
Arranging terms in Eq.(\ref{Eqmov}) we find the following differential equation
\begin{equation}
\frac{du(t)}{dt}= \alpha \Delta T_g g - \frac{6\pi D \mu (\phi r_F)^{1-D}}{\Omega_D \rho}u(t),
\end{equation}
where we have expressed the intruder's radius as function of the size ratio dependence
$r_I=\phi r_F$, and the solution of this differential equation is the rise velocity of
the intruder
\begin{equation}
u(t) = \frac{\alpha \Delta T_g g t_0}{\phi^{1-D}} \left[  1- \exp \left( -\phi^{1-D}\frac{t}{t_0} \right ) \right ],
\label{u}
\end{equation}
where the time--scale $t_0$ is given by
\begin{equation}
t_0 \equiv \frac{\Omega_D \rho }{6\pi D \mu r_F^{1-D}}.
\end{equation}

The force balance between the drag force ${\bf f}_d$ and the buoyant force ${\bf f}_b$ gives
the settling velocity $u_s$
\begin{equation}
u_s = \frac{\alpha  \Delta T_g t_0}{\phi^{1-D}}.
\label{us}
\end{equation}

The time dependent intruder height $z(t)$ is
\begin{equation}
z(t) = \frac{\alpha\Delta T_g g t_0}{\phi^{1-D}}\left[ t - t_0\phi \left(1-\exp\left(-\phi^{1-D}\frac{t}{t_0}  \right)\right) \right].
\label{z}
\end{equation}

On a qualitative level our model satisfactorily reproduces the observed phenomenology:
a large intruder migrates to the top of a vibrated bed, and the rise velocity increases
with the intruder size. The solutions (\ref{u}) and (\ref{z}) are plotted in Figs. 3 and 4.
Our results resembles the experimental intruder height time evolution described in Refs.\cite{Duran94}
and \cite{Cooke96}. However the model can not describe the intermittent ascent of the
intruder since we calculate the mean velocities.
Using the following model parameters: mass particle density of 2.7 gcm$^{-3}$
(Aluminum), $r_F=0.1$ cm, $e=0.83$, $\nu=0.34$, $N= 5\times 10^3$, $g=100$ cms$^{-2}$, $r_0=2r_I$ and
$T_0=30\times 10^{6}$ gcm$^2$s$^{-2}$ we obtain that the order of magnitude of $z(t)$ (Fig. 4) coincides with
the values reported by Cooke {\it et al.} (See Fig. 3 Ref.\cite{Cooke96}).

\begin{figure}[t]
\centerline{\psfig{file=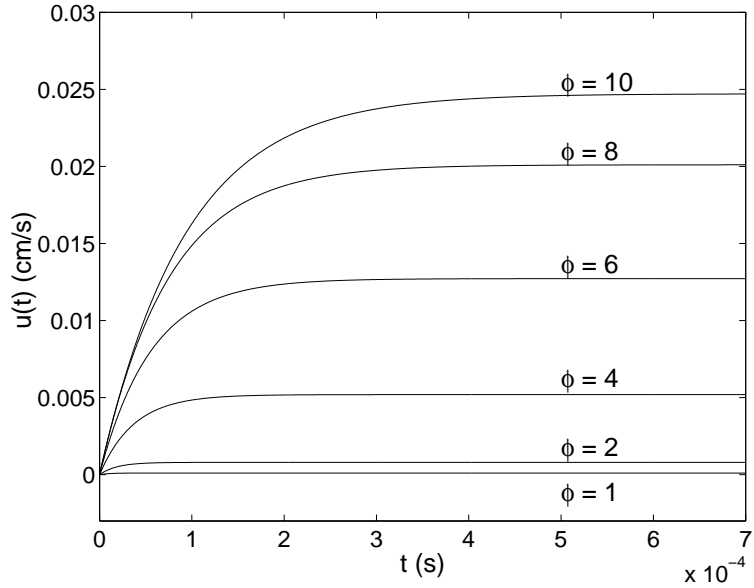,width=10.0cm}}
\caption{Intruder segregation velocity $u(t)$. The  parameters are:
mass particle density of 2.7 gcm$^{-3}$
(Aluminum), $r_F=0.1$ cm, $e=0.83$, $\nu=0.34$, $N= 5\times 10^3$, $g=100$ cms$^{-2}$, $r_0=2r_I$ and
$T_0=30\times 10^{6}$ gcm$^2$s$^{-2}$.
\label{fig3}}
\end{figure}

\begin{figure}[t]
\centerline{\psfig{file=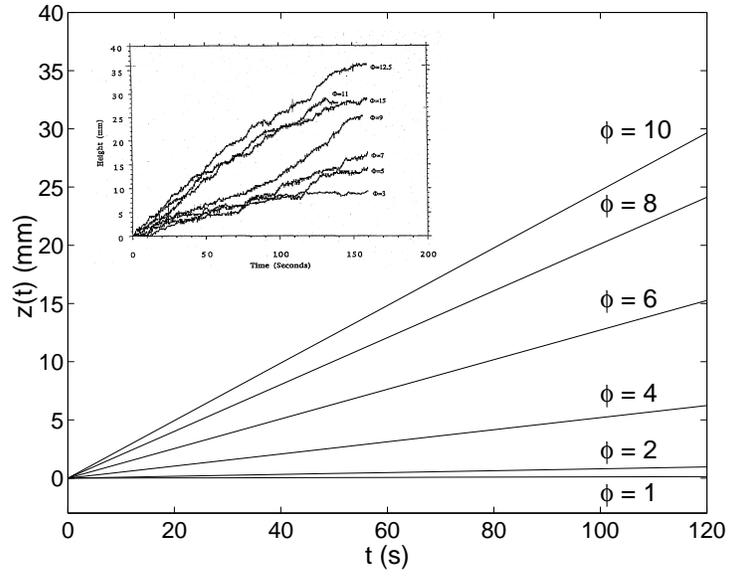,width=15.0cm}}
\caption{Intruder height time dependence $z(t)$. The  parameters are:
mass particle density of 2.7 gcm$^{-3}$
(Aluminum), $r_F=0.1$ cm, $e=0.83$, $\nu=0.34$, $N= 5\times 10^3$, $g=100$ cms$^{-2}$, $r_0=2r_I$ and
$T_0=30\times 10^{6}$ gcm$^2$s$^{-2}$.
Inset: Measured intruder height (Fig. 3, Ref.[6]).
\label{fig4}}
\end{figure}

From the settling velocity $u_s$ (\ref{us}) we show explicitly the
dependence on size. This is proportional to the size ratio $\phi$ and the granular temperature
differences $\Delta T_g$ which also depends on the size ratio. Its agrees with the
experimental fact that the larger the radius of the intruder, the faster is the ascent, reported
by Duran {\it et al.}\cite{Duran94}. The plotted solution (\ref{us}), Fig. 5, fits well with the
experimental values from Ref.\cite{Duran94} for $\phi>4$, shown in Fig. 5. In this experiment, Duran {\it et al.,}
claim the experimental evidence of a segregation size threshold at $\phi_c=3.3$,
below which the intruder does not
exhibit any upwards motion. Our model's continuous aspect doesn't allow for the existence of this
threshold. We argue that this discrepancy comes from the fact that  experimental measures in this
regime should be very difficult to carring out.
 They reported to not have observed any upward motion after one hour. If we see the
height profile shown in Fig. 4 for $\phi=2$, we note that it is very close to zero after one minute and
the slope is very low. Therefore this threshold should not exist.

\begin{figure}[t]
\centerline{\psfig{file=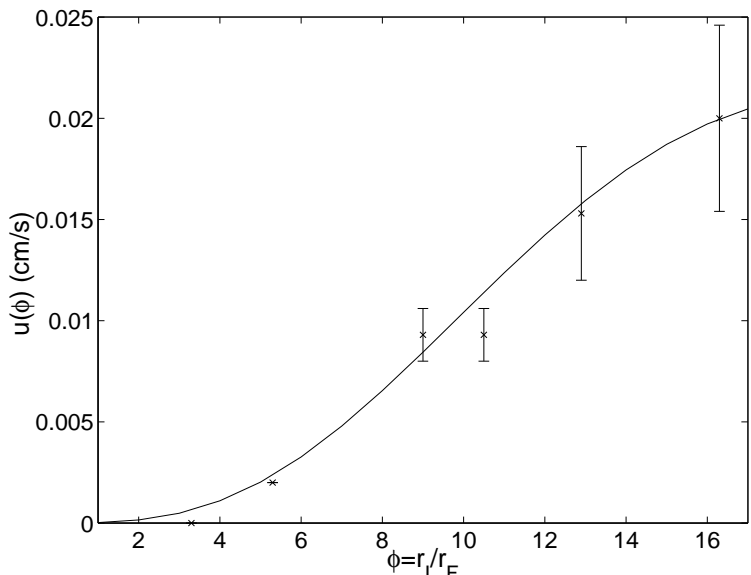,width=10.0cm}}
\caption{Intruder segregation velocity dependence on $\phi$. The  parameters are:
mass particle density of 2.7 gcm$^{-3}$
(Aluminum), $r_F=0.75$ cm, $e=0.85$, $\nu=0.25$, $N= 5\times 10^3$, $g=100$ cms$^{-2}$, $r_0=2r_I$ and
$T_0=126\times 10^{6}$ gcm$^2$s$^{-2}$. The data points come from Ref.[5].}
\end{figure}

\section{Conclusions}

We derived a phenomenological continuum description for particle size segregation in granular media.
We propose a buoyancy--driven segregation mechanism caused by the dissipative nature of the collisions
between grains. The collisional dissipation rate naturally leads  to a local temperature difference among
the region around the intruder and the medium without intruder. In this model we proposed that the
intruder's presence develops a temperature gradient in the system which gives origin to a difference of
densities. The granular temperature difference is due to the fact that the number of collisions on the
surface increases with the size of the intruder, but the local density of dissipated energy diminishes.
So, the region around the intruder its {\it hotter} than the region without intruder. From
this temperature difference we can conclude that we have a change in the {\it effective density}
of the granular fluid. This leads to an effective buoyancy force that is the responsible for
the intruder's upward movement.

In this work we made use of the tools of  kinetic theory of gases to calculate the granular temperature.
We observed a breakdown of the energy equipartition. And this is in agreement with other reported
experiments and models. In a certain sense our theory unifies the different aspects observed in the size
segregation phenomenon. Explicit solutions of the dependence of height and velocity are calculated.
The geometrical effect of a segregation threshold is not supported by our model. The intruder size
dependence appears naturally in our model.

Very recently it has been shown experimentally \cite{Wildman01} and by computer simulations\cite{Ramirez00},
that the convection phenomenon in granular fluids comes from the effect of spontaneous granular temperature
gradients, due to the dissipative nature of the collisions. This temperature gradient leads to a density
variations. The convection rolls are caused by buoyancy effects initiated by enhanced dissipation at the
walls and the tendency of the grains at the center to rise. So, this segregation mechanism could be described
in the hydrodynamic framework proposed in this work subject to the appropriate boundary conditions.

\begin{acknowledgments}

We thank M. Alam for helpful comments on the manuscript. One of the authors
(L.T.) would like to thank A.R. Lima for friendly support.
\end{acknowledgments}

\appendix
\section{Transport coefficients}

In this appendix the prefactors appearing in Eqs.(\ref{conductivity}) and (\ref{viscosity})
are derived. Using a
Chapman--Enskog procedure for the solution of the Enskog transport equation, the
transport coefficients for nearly elastic particles have been derived in Refs.\cite{Jenkins85}
and \cite{Jenkins98}.

In $2D$ the thermal conductivity $\kappa$ is\cite{Jenkins85}
\begin{equation}
\kappa = 3n r_F\left( \frac{\pi}{m_F} \right)^{1/2}\left[ 1 + \frac{1}{3}\frac{1}{G} + \frac{3}{4} \left( 1+\frac{16}{9\pi} \right)G\right] T_g^{1/2},
\label{k1}
\end{equation}
where $G$ is $\nu g_0$,
$g_0$ is the $2D$ pair correlation function given in Eq.(\ref{Verlet}), and
$\nu$ is the area fraction $\nu = n\pi r_F^2$.
It is convenient to express Eq.(\ref{k1}) introducing the prefactor $\kappa_0$ deffined as
\begin{equation}
\kappa_0 \equiv 3n r_F\left( \frac{\pi}{m_F} \right)^{1/2}\left[ 1 + \frac{1}{3}\frac{1}{G} + \frac{3}{4} \left( 1+\frac{16}{9\pi} \right)G\right].
\end{equation}
The result (\ref{k1}) takes the form
\begin{equation}
\kappa = \kappa_0\sqrt{T_g}.
\end{equation}

In $3D$ the thermal conductivity is\cite{Jenkins98}
\begin{equation}
\kappa = \frac{15}{8}n r_F \left( \frac{\pi}{m_F} \right)^{1/2}\left[ 1 + \frac{5}{24}\frac{1}{G}+\frac{6}{5}\left( 1 + \frac{32}{9\pi} \right) G \right] T_g^{1/2},
\end{equation}
where $G$ is $\nu g_0$,
$g_0$ is the $3D$ pair correlation function given in Eq.(\ref{Carnahan}), and
$\nu$ is in this case the volume fraction $\nu = 4\pi n r_F^3/3$.
In $3D$ the prefactor $\kappa_0$ is defined as
\begin{equation}
\kappa_0 \equiv \frac{15}{8}n r_F\left( \frac{\pi}{m_F} \right)^{1/2}\left[ 1 + \frac{5}{24}\frac{1}{G}+\frac{6}{5}\left( 1 + \frac{32}{9\pi} \right) G \right].
\end{equation}

The shear viscosity $\mu$ in $2D$ is\cite{Jenkins85}

\begin{equation}
\mu = \frac{1}{4} n r_F (\pi m_F)^{1/2} \left[ 2 + \frac{1}{G} + \left( 1 + \frac{8}{\pi} \right)G \right]T_g^{1/2}.
\label{mu2D}
\end{equation}

It is convenient to express Eq.(\ref{mu2D}) introducing the prefactor $\mu_0$ defined as
\begin{equation}
\mu_0 = \frac{1}{4} n r_F (\pi m_F)^{1/2} \left[ 2 + \frac{1}{G} + \left( 1 + \frac{8}{\pi} \right)G \right].
\end{equation}

So, the result (\ref{mu2D}) takes the form
\begin{equation}
\mu = \mu_0\sqrt{T_g}.
\end{equation}

In  $3D$ the shear viscosity is\cite{Jenkins98}
\begin{equation}
\mu = \frac{1}{3}nr_F(\pi m_F)^{1/2}\left[ 1 + \frac{5}{16}\frac{1}{G} + \frac{4}{5}\left( 1 + \frac{12}{\pi} \right)G \right]T_g^{1/2},
\end{equation}
and the prefactor $\mu_0$ in $3D$ is defined as
\begin{equation}
\mu_0 = \frac{1}{3}nr_F(\pi m_F)^{1/2}\left[ 1 + \frac{5}{16}\frac{1}{G} + \frac{4}{5}\left( 1 + \frac{12}{\pi} \right)G \right].
\end{equation}

\section{Thermal expansion coefficient}

We can consider the volume of the system as a function of the granular temperature and the
pressure $V=V(T_g,p)$. A change in the granular temperature $dT_g$ and the pressure
$dp$, leads to the corresponding change in the volume $dV$
\begin{equation}
dV =\left(\frac{\partial V}{\partial T_g}\right)_p dT_g + \left(\frac{\partial V}{\partial p}\right)_{T_g} dp.
\end{equation}
As we have supposed that the pressuer of the system is more or less constant, we can
approximate $dp\sim 0$. The increment of volume $dV$ with an increment of the granular
temperature $dT_g$ is
\begin{equation}
dV =\left(\frac{\partial V}{\partial T_g}\right)_p dT_g.
\end{equation}
Thus,
\begin{equation}
\frac{dV}{dT_g} =\left(\frac{\partial V}{\partial T_g}\right)_p,
\end{equation}
or
\begin{equation}
\left(\frac{\partial V}{\partial T_g}\right)_p = \left[ \left(\frac{\partial T_g}{\partial V}\right)_p \right]^{-1},
\end{equation}
and in terms of the number density $n$, we have
\begin{equation}
\left(\frac{\partial n}{\partial T_g}\right)_{p,N} = \left[ \left(\frac{\partial T_g}{\partial n}\right)_{p,N} \right]^{-1}.
\label{lemma}
\end{equation}
From the definition of the coefficient of thermal expansion Eq.(\ref{alfa}), and from the
above statement, we find
\begin{equation}
\alpha = -\frac{1}{n}\left( \frac{\partial n}{\partial T_g} \right)_{p,N} = \left[ \left( \frac{\partial T_g}{\partial n} \right)_{p,N} \right]^{-1}.
\label{alfa2}
\end{equation}
The partial derivative $(\partial T_g/\partial n)_{p,N}$ can be calculated from the equation of
state (\ref{eqstate2}). In $2D$ the equation of state is
\begin{equation}
p= \frac{1+e}{2}nT_g\left[ 1 + 2\nu \frac{\left( 1-\frac{7}{16}\nu \right) }{\left( 1-\nu \right)^2 } \right],
\label{eqstate2}
\end{equation}
where $\nu= n\pi r_F^2$. So, an elementary calculation leads to
\begin{equation}
\left( \frac{\partial T_g}{\partial n} \right)_{p,N}=-\frac{2}{(1+e)}\frac{p}{n^2}\frac{8(\nu^3-3\nu^2-8\nu -8)(\nu -1)}{(\nu^2+8)^2}.
\end{equation}
From Eq.(\ref{alfa2}) one obtains:
\begin{equation}
\alpha = \frac{2}{1+e}\frac{n}{p}\frac{(\nu^2+8)^2}{8(\nu^3-3\nu^2-8\nu -8)(\nu -1)}.
\end{equation}
Using the equation of state (\ref{eqstate2}) we can express $\alpha$ in function of the
granular temperature
\begin{equation}
\alpha = \frac{1}{T_g}\frac{(\nu^2+8)^2}{(\nu^3-3\nu^2-8\nu -8)(\nu -1)},
\end{equation}
this is
\begin{equation}
\alpha = \frac{1}{T_g}C(\nu),
\end{equation}
where the correction coefficient due to the density of the system is defined as
\begin{equation}
C(\nu)\equiv \frac{(\nu^2+8)^2}{(\nu^3-3\nu^2-8\nu-8)(\nu -1)}.
\end{equation}

For three dimensions the equation of state is
\begin{equation}
p=\frac{1+e}{2} nT_g \left[ 1 + 4\nu \frac{(2-\nu)}{2(1-\nu^3)} \right].
\end{equation}
In a similar way we find for $3D$ that the coefficient of thermal expansion is
\begin{equation}
\alpha = \frac{1}{T_g}\frac{(\nu^3 -\nu^2-\nu -1)(\nu -1)}{(\nu^4-4\nu^3 +4\nu^2+4\nu+1)},
\end{equation}
and the correction coefficient $C(\nu)$ in $3D$ is defined as
\begin{equation}
C(\nu)\equiv \frac{(\nu^3 -\nu^2-\nu -1)(\nu -1)}{(\nu^4-4\nu^3 +4\nu^2+4\nu+1)},
\end{equation}
where $\nu = 4n\pi r_F^3/3$.


\end{document}